\documentclass[11pt]{article}
\begin{document}
\title{\bf  Spin Particle with a Color Charge\newline
in a Color Field in Riemann--Cartan Space}
\author{O. V. Babourova,\thanks{E-mail:baburova@orc.ru}\\
Department of Theoretical Physics, Faculty of Physics,\\
Moscow State University, Leninskie Gory, d.1, st.2, Moscow 119992, Russia,\\
A. S. Vshivtsev,\thanks{Deceased}\\
Academician V. P. Myasnikov,\\
Keldysh Institute of Applied Mathematics, Russian Academy of Sciences,\\
Miusskaya pl. 4, Moscow 125045, Russia,\\
\and B. N. Frolov,\thanks{E-mail:frolovbn@orc.ru}\\
Department of Mathematics, Moscow State Pedagogical University,\\
Krasnoprudnaya 14, Moscow 107140, Russia
}
\date{}
\maketitle
\vskip 0.4cm
\par
{\bf Abstract—--}{\small On the basis of the method of Cartan exterior forms
and extended Lie derivatives, a hydrodynamic equation of the Euler type that
describes a perfect spin fluid with an intrinsic color charge in an external
non-Abelian color field in Riemann--Cartan space is derived from the energy-momentum
quasiconservation law. This equation is used to obtain a self-consistent
set of equations of motion for a classical test particle with a spin and a color
charge in a color field combined with a gravitational field characterized by curvature
and torsion. The resulting equations generalize the Wong equation, which describes
the motion of a particle with an isospin, and the Tamm--Good and Bargmann--Michel--Telegdi
equations, which describe the evolution of a charged-particle spin in an electromagnetic field.}
\vskip 0.4cm
\par

\section{INTRODUCTION}
\par
    In our previous study \cite{YAF}, we developed a variational theory of a perfect spin fluid
with an intrinsic non-Abelian color charge in Riemann--Cartan space $U_4$ with curvature and torsion.
This theory takes into account spin-polarization phenomena and chromomagnetic effects. In particular,
we obtained the equations of motion for such a fluid and the rules that govern the evolution
of a color charge and of a spin tensor satisfying the Frenkel condition. We also derived the
expression for the energy-momentum tensor of the fluid. The objective of this study is to deduce the
equation of motion for our color fluid in the form of a generalized hydrodynamic
equation of the Euler type in $U_4$ space and to obtain implications of this equation.
\par
    It is common knowledge that, in the theory of electromagnetic fields, the equations of motion of charged
particles follow from the law of energy-momentum conservation for the "particles + field" system and from
field equations, the law of energy-momentum conservation being a consequence of the invariance of the physical
system in question \cite{LL}, \cite{Br}. In the general theory of relativity, the equations of motion of
matter also follow from the equations for the  gravitational field. The reason is that the Einstein equations
imply the covariant conservation law for the matter energy-momentum. Conceptually, the same is true for field
theory in $U_{4}$ space, but the conservation laws have a more complicated structure in that case \cite{Tr}.
It was shown in \cite{BPF} that, in field theories formulated in spaces with a more complicated structure
($U_4$ space with Lagrangians quadratic in curvature and torsion and affine metric space with curvature, torsion,
and nonmetricity), the equations of motion for matter are obtained from  field equations and conservation laws
in just the same way.
\par
    This article is organized as follows. In Section 2, the method for obtaining the equations of motion
from conservation laws is used to derive the equation of motion for a perfect spin fluid having an intrinsic
color charge and interacting with a color field and with a gravitational field associated with the geometry
of $U_4$ space. We employ the method of Cartan exterior forms and extended Lie derivatives that take into account
internal gauge symmetries. In  \cite{prep}, this problem was solved by a different method. As a limiting case
of the equation obtained in Section 2, the equation of motion of a particle with a spin and with a non-Abelian
color charge in an external color field in $U_4$ space is derived in Section 3. This equation, which takes
into account spin-chromomagnetic effects, generalizes the well-known Wong equation \cite{Wo} of motion of
a particle with an isospin in a Yang--Mills field. In Section 4, the equation from Section 3 is used to obtain
the equation that describes the motion of a colored-particle spin in $U_4$ space and which generalizes the
Bargmann--Michel--Telegdi and Tamm--Good equations \cite{BMT}, \cite{BB}, \cite{prep}.
\par
    In this study, we adopt the notation and conventions introduced in \cite{YAF}.

\section{HYDRODYNAMIC EQUATION \newline OF MOTION FOR A SPIN FLUID \newline
         WITH AN INTRINSIC COLOR CHARGE}
\setcounter{equation}{0}
\par
    In Riemann--Cartan $U_4$ space, we consider a perfect spin fluid with a non-Abelian color charge in a color field.
The Lagrangian density of this system is given by
\begin{equation}
{\cal L}_{matter} = {\cal L}_{fluid} + {\cal L}_{field} \;, \label{eq:Lm}
\end{equation}
where the Lagrangian densities 4-forms ${\cal L}_{fluid}$ and ${\cal L}_{field}$ are presented in \cite{YAF}.
The invariance of this Lagrangian under general coordinate transformations and under local Lorentz transformations
implies fulfillment of the set of differential identities \cite{Tr}
\begin{eqnarray}
&&{\cal D}\Sigma_{a} = (\vec{e}_{a}\rfloor {\cal T}^{b})\wedge \Sigma_{b} - (\vec{e}_{a}
\rfloor {\cal R}^{c}\!_{b})\wedge \Delta^{b}\!_{c}\; , \label{eq:zak1}\\
&&{\cal D}\Delta_{ab} = - \theta_{[a}\wedge \Sigma_{b]}\; , \label{eq:zak2}
\end{eqnarray}
where $\wedge$ denotes exterior multiplication; $\rfloor$ denotes inner multiplication (contraction); $\theta^{a}$
are basis 1 -forms; ${\cal D}$ represents exterior covariant differentiation; ${\cal T}^{b}$ is the 2-form of torsion;
${\cal R}^{c}\!_{b}$ is the 2-form of the curvature of $U_{4}$ space; $\Sigma_{a}$ is the 3-form of the matter
energy-momentum; and${\Delta}_{ab}$ is the 3-form of the spin. Equation (\ref{eq:zak1}) follows from the invariance
of the Lagrangian under spacetime diffeomorphisms -- it generalizes the law of energy-momentum conservation --
while equation (\ref{eq:zak2}) appears to be a consequence of the invariance of the Lagrangian under local Lorentz
transformations, generalizing the law of spin-momentum conservation.
\par
    If the equations of motion for the physical system being considered are satisfied in $U_{4}$ space, equations
(\ref{eq:zak1}) and (\ref{eq:zak2}) hold identically. As a matter of fact, they offer an alternative representation
of the equations of motion. We will now use relation (\ref{eq:zak1}) to obtain the equation that describes the motion
of a perfect spin fluid with an intrinsic color charge and which has a form generalizing the hydrodynamic Euler equation.
To do this, the expression obtained in \cite{YAF} for the 3-form of the energy-momentum of the system in question
\{equation (5.9) in [1]\} is substituted into equation (\ref{eq:zak1}). This 3-form can be represented as
\begin{eqnarray}
\Sigma_{a} &=& p(\vec{e}_{a}\rfloor \eta) + n\left (\pi_{a} +\frac{p}{nc^{2}}u_{a}
\right ) u + \chi n(\vec{e}_{a}\rfloor {\cal F}^{m})\wedge J_{m}*\!{\cal S} +
\nonumber \\
&& + \frac{\alpha}{2} \left ((\vec{e}_{a}\rfloor {\cal F}^{m})\wedge *\!{\cal F}_{m}
- {\cal F}^{m}\wedge (\vec{e}_a \rfloor *\!{\cal F}_m) \right ) \; , \label{eq:sig}
\end{eqnarray}
where $\eta$ is the 4-form of volume; $n$ is the particle-number density of the fluid; $u$ is the 3-form
of the particle velocity; $J_m$ is the color charge of a fluid element \{see equation (2.4) in \cite{YAF}\};
${\cal S} = (1/2) S_{ab}\theta^a \wedge\theta^b$ is the 2-form of the spin of a fluid element ($S_{ab}$
is the spin tensor); and $\pi_a$ is the dynamical momentum of a fluid element. According to equation (5.7)
from \cite{YAF}, this momentum is given by
\begin{eqnarray}
&&\pi_{a} = \frac{1}{c^{2}}\varepsilon^{*}u_{a} - \frac{1}{c^{2}}S_{a}^{~c}
\left (\dot{u}_{c} + u^{b}(\chi J_{m}F^{m}\!_{bc} + \omega_{bc})\right ) \; ,
\label{eq:pi} \\
&&\varepsilon^{*}\eta = \frac{\varepsilon}{n}\eta + \chi J_{m}{\cal F}^{m}
\wedge *{\cal S}\; , \label{eq:ef}
\end{eqnarray}
where $\varepsilon$ is the energy density of the fluid in the reference frame comoving with it, while
$\varepsilon^{*}$ is the effective energy per a particle of the fluid \cite{prep}. In these formulas,
an overdot denotes differentiation defined for an arbitrary tensor $\Phi$ as
\begin{equation}
\dot{\Phi}^{a}\!_{b} := *\!(u\wedge {\cal D}\Phi^{a}\!_{b})\; .\label{eq:27}
\end{equation}
\par
    In order that the Bianchi identities ${\cal D}{\cal F}^{m} = 0$ for the 2-form of the strength tensor
of the color field could be used explicitly, the operation of exterior covariant differentiation ${\cal D}$ must
be defined with respect to the two gauge groups [Lorentz group and $SU(3)$ color group] according to equation (3.2)
from [1]. In calculating covariant derivatives of contractions, this implies the use of the Lie derivative
$\pounds_{{\small \vec{v}}}$ generalized to arbitrary local gauge groups [including the $SU(3)$ color group];
that is,
\begin{equation}
{\cal D}\circ i_{{\small \vec{v}}} = \pounds_{{\small \vec{v}}}
- i_{{\small \vec{v}}} \circ {\cal D}\; ,\label{eq:li}
\end{equation}
where  $i_{{\small\vec{v}}}$  is the operator of inner multiplication with respect to the vector
$\vec{v}$. The calculations must rely on the continuity equation $d(nu) =  0$; on the relation
\begin{equation}
{\cal D}(\chi n J_{m}*\! {\cal S}) = n J_{m} u -
{\cal D}(\alpha *\!{\cal F}_{m})\; , \label{eq:A}
\end{equation}
which follows from the equation for the color gauge field \{see equation (4.3) in [1]\}; and on the
equations
\begin{equation}
{\cal D}(\vec{e}_a \rfloor \eta) = {\cal D}\eta_a = (\vec{e}_a \rfloor {\cal T}^{b})\wedge
\eta_b \;. \label{eq:eta}
\end{equation}
which are valid in $U_4$ space.
\par
    In the calculations, there arises the differential operator
\begin{equation}
\pounds_{\vec{e}_a}{\cal F}^m - (\vec{e}_a \rfloor {\cal T}^b) \wedge (\vec{e}_b \rfloor
{\cal F}^m) =: (\vec{e}_a \rfloor \nabla) {\cal F}^m \;, \label{eq:dif}
\end{equation}
which represents the total covariant derivative of tensor-valued forms and which takes into account
all the types of indices [spinorial and tensorial indices associated with the representations of the
Lorentz and $SU(3)$ groups, as well as indices associated with the space of $p$-forms]. It can be
shown that the operator in (\ref{eq:dif}) satisfies the condition
\begin{equation}
*\! (\vec{e}_a \rfloor \nabla) {\cal F}^m = (\vec{e}_a \rfloor \nabla) *\!{\cal F}^m \;.
\label{eq:st}\end{equation}
\par
    Substituting (\ref{eq:sig}) into (\ref{eq:zak1}), using equations (\ref{eq:li})--(\ref{eq:st}),
and considering that \{see equation (5.1) in [1]\} the expression for the 3-form of the spin momentum is
\begin{equation}
{\Delta}_{ab} = \frac{1}{2} S_{ab} u \; ,
\end{equation}
we arrive at the equation of motion for the fluid in the form
\begin{eqnarray}
u\wedge {\cal D}\left (\pi_{a} + \frac{p}{nc^{2}}u_{a}\right ) = \frac{1}{n}
\eta (\vec{e}_a \rfloor \nabla) p - (\vec{e}_{a}\rfloor {\cal T}^{b})\wedge \left ( \pi_{b} +
\frac{p}{nc^{2}}u_{b} \right ) u - \nonumber \\
- \frac{1}{2} (\vec{e}_{a}\rfloor {\cal R}^{bc})\wedge S_{bc}u -
(\vec{e}_{a}\rfloor {\cal F}^{m})\wedge J_{m}u +
\chi (\vec{e}_a \rfloor \nabla) {\cal F}^m \wedge J_{m} *\!{\cal S} \; ,
\label{eq:euler}
\end{eqnarray}
which represents a generalization of the well-known hydrodynamic Euler equation to the case of
a perfect spin fluid with a color charge.

\section{EQUATIONS OF MOTION OF A PARTICLE \newline WITH A SPIN AND A COLOR CHARGE}
\setcounter{equation}{0}
\par
    By going over to the limit of zero pressure in equation (\ref{eq:euler}), we find that, in $U_4$ space,
the equation of motion of a particle with a spin and a color charge in an external non-Abelian color gauge
field has the form
\begin{eqnarray}
&& u\wedge {\cal D}\pi_{a} = - (\vec{e}_{a}\rfloor {\cal F}^{m})\wedge J_{m}u
+ \chi (\vec{e}_a \rfloor \nabla) {\cal F}^m \wedge J_{m} *\!{\cal S} -
\nonumber \\
&& - \frac{1}{2} (\vec{e}_{a}\rfloor {\cal R}^{bc})\wedge S_{bc}u
- (\vec{e}_{a}\rfloor {\cal T}^{b})\wedge \pi_{b}\; . \label{eq:part}
\end{eqnarray}
The first term on the right-hand side of this equation is a generalization of the Lorentz force to the case
of a non-Abelian gauge field. The second term is a chromo-magnetic analog of the Stern--Gerlach force acting
on a magnetic moment in an electromagnetic field (this force is generated by the additional potential energy
of a magnetic moment in a magnetic field \cite{BB}). The third term on the right-hand side of equation
(\ref{eq:part}) represents the Mathisson force arising from the interaction of the particle spin with the
curvature of space, while the fourth term is the so-called translational force, which is due to the interaction
of the particle dynamical momentum with the torsion of space. The emergence of this force is peculiar to $U_{4}$ space.
\par
    Equation (\ref{eq:part}) must be supplemented with the law that governs spin evolution
\{see equation (3.18) in [1]\},
\begin{equation}
u\wedge {\cal D}S_{ab} = 2 \pi_{[a}u_{b]}\eta - 2 S_{[a}^{~c}
(\chi \eta_{b]c}\wedge {\cal F}^{m}J_{m} + \omega_{b]c}\eta) \; ,
\label{eq:spin}\end{equation}
and with the law that governs the evolution of the particle color charge \{see equation (4.4) in [1]\},
\begin{equation}
u\wedge {\cal D}J_{m} = - c_{m}\!^{p}\!_{n} J_{p} (\chi {\cal F}^{n}\wedge
*{\cal S} + \omega^{n} \eta )\; , \label{eq:J}
\end{equation}
where $\eta_{bc} = \vec{e}_c \rfloor \eta_b = \vec{e}_c \rfloor \vec{e}_b \rfloor\eta = *(\theta_b
\wedge \theta_c)$, and $c_{m}\!^{p}\!_{n}$ are the structure constants of the $SU(3)$ group.
\par
    Equations (\ref{eq:part})--(\ref{eq:J}) describe the motion of a particle with a spin and a color charge
in the presence of spin-chromomagnetic interaction. They generalize the well-known Wong equations \cite{Wo}
to the case of the $SU(3)$ color group and take into account the particle spin, which may be responsible
for the possible interaction between the spin and the chromomagnetic component of the color field and for
the additional effect of the gravitational field associated with the geometry of Riemann--Cartan space
on the motion of the particle being considered.

\section{EVOLUTION OF THE PARTICLE SPIN IN \newline
         A COLOR FIELD IN RIEMANN--CARTAN SPACE}
\setcounter{equation}{0}
\par
    The particle-spin vector (Tamm--Pauli--Lyubanski vector) is defined as
\begin{equation}
\sigma^a := \frac{1}{2c^2}\eta^{abcd}S_{bc}u_d \;, \qquad \sigma :=
\sqrt{\sigma^a \sigma_a} \label{eq:paul}
\end{equation}
where $\eta^{abcd}$ are the components of the Levi-Civita antisymmetric tensor (4-form of volume $\eta$).
In Minkowski spacetime, the evolution of this vector in a slowly varying external electromagnetic field
is governed by the Bargmann--Michel--Telegdi equation \cite{BMT}. However, this equation takes no account
of the effect of the spin on the trajectory of the particle. A more precise equation that describes
the motion of the particle spin in a non-uniform electromagnetic field and which takes into account
the effect of the spin on the motion of the particle was derived by Good who extended
the Tamm equation (see \cite{BMT}, \cite{BB}).
\par
    By using equations (\ref{eq:part}) and (\ref{eq:spin}), which describe the motion of a particle
with a spin and a color charge, we will now extend the Tamm--Good and Bargmann--Michel--Telegdi equations
to the case of the motion of such a particle in an external color (generally nonuniform) field in
$U_4$ Riemann--Cartan space \cite{prep}. To this end, we recast the dynamical momentum of the particle
[it is given by (\ref{eq:pi})] into the form
\begin{eqnarray}
&& \pi_a = m^{*}u_a - \lambda_a \;, \qquad m^{*} := \frac{\varepsilon^{*}}
{c^2} \;, \label{eq:pi1} \\&&\lambda_a :=\frac{1}{c^2} S_a\!^c w_c \;,\qquad
w_c := \dot{u}_c + u^b (\chi J_m F^m\! _{bc} + \omega_{bc})\; . \label{eq:pi2}
\end{eqnarray}
Using definitions (\ref{eq:paul}) and (\ref{eq:pi2}) and considering that the spin tensor,
which is the inverse of that in (\ref{eq:paul}), is given by
\begin{equation}
S_{ab} = \frac{1}{c}\eta_{abcd}\sigma^{c}u^{d}\; , \label{eq:tens}
\end{equation}
we can easily prove that the vectors $u^a$, $\sigma^a$, and $\lambda^a$ are mutually orthogonal.
\par
    Differentiating the spin vector (\ref{eq:paul}), we arrive at
\begin{equation}
u\wedge{\cal D}\sigma^a = \frac{1}{m^{*}c^2}u^a\sigma^b
u\wedge {\cal D} (\pi_b - \lambda_b ) + \sigma^b\Pi^{ac}(\chi\eta_{bc}\wedge
{\cal F}^m J_m + \omega_{bc} \eta)\;, \label{eq:sig1}
\end{equation}
where $\Pi^{ac} = g^{ac} + (1/2c^2) u^a u^c$ is the projection tensor. To derive this relation,
we used equation (\ref{eq:spin}), which governs the evolution of the spin tensor; the Frenkel
condition $S_{ab}u^b  = 0$; expression (\ref{eq:tens}) for the spin tensor in terms of the spin vector;
the prescription $\eta_{abcd}\eta^{ijkd}= -6\delta^i_{[a}\delta^j_b\delta^k_{c]}$ for contracting
indices of the Levi--Civita tensor; and the result obtained by differentiating expression (\ref{eq:pi1})
for the dynamical momentum,
\begin{equation}
m^{*}u\wedge{\cal D} u_a =  u\wedge {\cal D} (\pi_a + \lambda_a )
+ u_a u\wedge{\cal D}m^{*} \; . \label{eq:sig2}
\end{equation}
Differentiating expression (\ref{eq:pi2}) for $\lambda_a$ and taking into account equation (\ref{eq:spin})
and the orthogonality conditions $\sigma^a u_a = 0$,  $\sigma^a \pi_a = 0$ and $\sigma^a S_{ab}= 0$,
we obtain
\begin{equation}
\sigma^a u\wedge{\cal D}\lambda_a =
\lambda^a\sigma^b (\chi\eta_{ab}\wedge {\cal F}^m J_m + \omega_{ab}\eta)\; .
\end{equation}
Substituting this equation and the equation (\ref{eq:part}) of the motion of a particle into
(\ref{eq:sig1}) and taking into account equation (\ref{eq:pi1}), we arrive at
\begin{eqnarray}
&&u\wedge {\cal D}\sigma^{a} = - \sigma^b (\chi\eta^{a}_{~b}\wedge {\cal F}^m
J_m +\omega^{a}_{~b}\eta ) - \nonumber \\
&&- \frac{u^{a} \sigma^{b}}{m^{*}c^{2}}
[ (e_{b}\rfloor {\cal F}^{m})\wedge J_{m}u - \pi^c (\chi\eta_{bc}\wedge
{\cal F}^{m} J_m + \omega_{bc} \eta ) - \nonumber \\
&&- \chi (\vec{e}_b\rfloor\nabla){\cal F}^{m}\wedge J_{m} *\!{\cal S}
+ \frac{1}{2} (\vec{e}_{b}\rfloor {\cal R}^{cd})\wedge S_{cd}u
+ (\vec{e}_{b}\rfloor {\cal T}^{c})\wedge \pi_{c} u ]\; . \label{eq:BMT}
\end{eqnarray}
\par
    Equation (\ref{eq:BMT}) represents the sought extension of the Tamm--Good and Bargmann--Michel--Telegdi
equations to the case of a particle having a spin and a color charge and moving in an external
nonuniform non-Abelian color field in Riemann--Cartan space.

\section{CONCLUSION}
\par
    On the basis of the method of Cartan exterior forms and generalized Lie derivatives, a hydrodynamic equation
of the Euler type that describes the motion of an element of a perfect spin fluid with a color charge in a
non-Abelian color gauge field associated with the $SU(3)$ group and a gravitational field described by the
geometry of Riemann--Cartan space with curvature and torsion has been derived from the law of energy-momentum
quasiconservation. The "fluid + field" system being considered is closed because the gravitational field is
generated by the energy-momentum tensor of the fluid and because the color field is induced by the
current of non-Abelian color charge of the fluid.
\par
    From the resulting hydrodynamic equation, we have deduced an equation that describes the motion of
a particle with a spin and a color charge. We have shown that the forces acting on such a particle represent
generalizations of forces of well-known types. These are a Lorentz force acting on the non-Abelian color charge;
the Stern--Gerlach force, which is proportional to the gradient of the color field; the Mathisson force resulting
from the interaction between the particle spin and the curvature of spacetime; and the force of the translational
type caused by the interaction between the particle momentum and the torsion of spacetime. It is noteworthy that
forces analogous to the Mathisson force and to the translational force appear in modern gauge theories of plasticity
that employ gauge groups to describe defects in crystals, disclinations in the case of the gauge group of rotations
and dislocations in the case of the gauge group of translations \cite{KE}, \cite{GuM}.
\par
    Finally, the equation of motion of a particle has been used to deduce an equation that describes the evolution
of the particle-spin vector in external color and gravitational (with curvature and torsion) fields and which
generalizes, to this case, the known equations that describe motion of the vector of the charged-particle spin in
an electromagnetic field (Bargmann--Michel--Telegdi equation for a uniform field and Tamm--Good equation for
a nonuniform field) \cite{BMT}, \cite{BB}, \cite{prep}.
\par
    It is appropriate to compare our equations of motion of a classical particle and of a classical spin vector
with the equations representing the limiting case of the corresponding quantum equations obtained by using QED methods.
In QED, neither the Bargmann--Michel--Telegdi equation nor the equation of motion of a particle (Lorentz equation)
involves gradient terms in the original forms of these equations. That there are no such terms is due to the use
of the procedure that derives these equations for constant fields. Nevertheless, terms of this type arise
in the equations of motion if a consistent procedure based on the Maslov method \cite{VPM} and on its ensuing
development \cite {R} is invoked to deduce the classical equations of motion for particles and the equations
of motion for spin and isospin vectors from the quantum Dirac equation with allowance for external fields \cite{S},
\cite{T}. Therefore, these terms make it possible to demonstrate nontrivial quantum effects at the classical level.
\par
    Note that, apart from terms involving field gradients, our equations of motion for a fluid element coincide
with the equations of motion that were obtained by the path-integral method \cite {Shv}. They also coincide
in form with the corresponding equations of motion that follow from the string action functional \cite{Ir}.
This suggests that our equations of motion may prove to be valid not only for a fluid element but also for
extended objects like strings. This gives reason to hope that the Weyssenhoff--Raabe model of a fluid can be
extended to nonlocal objects and that some additional analogous structures can be recognized in doing this.
An important point is that the Weyssenhoff-Raabe model is a classical model of a fluid, but it describes the
properties of a second-quantized fermion field rather than of Dirac electrons associated with a nonquantized
spinor field, as is often erroneously assumed. This property of the Weyssenhoff--Raabe model stems from
the well-known fact that the energy of the spinor field becomes positive definite only after Fermi--Dirac
second quantization.
\par
    To develop this theory further, it is advisable to study the effect of a dilaton field on the motion
of a particle with a spin and a color charge. For this, a particle must be endowed with a dilaton charge,
and the theory developed in \cite{dil} to describe the motion of a particle with a spin
and a dilaton charge in Riemann--Cartan and Weyl--Cartan spaces must be taken into consideration.
A self-consistent theory constructed in this way can be used to study a nonperturbative gluon condensate
up to the point of the phase transition from hadron matter to quark-gluon plasma.

\end{document}